\documentclass[pra,aps,preprint,preprintnumbers,showpacs,amsmath,amssymb,secnumroman,eqsecnum]{revtex4}

\usepackage{amsmath}
\usepackage{amssymb}
\usepackage{graphicx}
\usepackage{dcolumn}
\usepackage{bm}

\newcommand{\beq}{\begin{equation}}
\newcommand{\eeq}{\end{equation}}
\newcommand{\beqa}{\begin{eqnarray}}
\newcommand{\eeqa}{\end{eqnarray}}

\newcommand{\vol}[1]{{\bf #1}}

\newcommand{\du}[1]{{\bf\sf #1}}

\begin{document}


\title{Radiation from a semi-infinite unflanged planar dielectric waveguide}

%

\author{B. U. Felderhof}

 \email{ufelder@physik.rwth-aachen.de}
\affiliation{Institut f\"ur Theorie der Statistischen Physik\\ RWTH Aachen University\\
Templergraben 55\\52056 Aachen\\ Germany\\
}%

\date{\today}

\begin{abstract}
Radiative emission from a semi-infinite unflanged planar dielectric waveguide is studied for the case of TM polarization on the basis of an iterative scheme. The first step of the scheme leads to approximate values for the reflection coefficients and electromagnetic fields inside and outside the waveguide. It is shown that for the related problems of reflection from a step potential in one-dimensional quantum mechanics and of Fresnel reflection of an electromagnetic plane wave from a half-space the iterative scheme is in accordance with the exact solution.
\end{abstract}

\maketitle

\section{\label{I} Introduction}

In a classic paper Levine and Schwinger \cite{1} studied the radiation of sound from an unflanged circular pipe. Later they extended their theory to electromagnetic radiation \cite{2}. Their work constituted the first major advance in the theory of diffraction after Sommerfeld's exact solution of the problem of plane wave diffraction by an ideally conducting half-plane \cite{3}. In the theory of diffraction of sound, microwaves, or light, the radiation is assumed to propagate in uniform space with reflection by rigid objects or idealized boundaries. For a lucid introduction to the theory of diffraction we refer to Sommerfeld's lecture notes \cite{4}. The early theory of diffraction was reviewed by Bouwkamp \cite{5}. Later developments are discussed by Born and Wolf \cite{6}. A brief review of the principles and applications of open-ended waveguides with idealized walls was presented by Gardiol \cite{7}.

The invention of the dielectric waveguide by Hondros and Debye \cite{8} has led to the development of optical fibers and the subsequent advances in telecommunication. In the theoretical determination of running wave solutions to Maxwell's equations in a spatially inhomogeneous medium the radiation is assumed to propagate in a guiding structure of infinite length. The problem of emergence of radiation from a semi-infinite waveguide into a half-space is of obvious technical interest. In the case of sound the analysis is based on the exact solution of Levine and Schwinger \cite{1},\cite{9,9A,10,11}. For a dielectric waveguide the observation of the emerging radiation can be used as a tool to study the nature of the driving incident wave \cite{12}.

It is advantageous to simplify the theoretical analysis by the use of planar symmetry. The theory of Levine and Schwinger was extended to planar geometry by Heins \cite{13,13A}. In the following we study radiation emerging from a semi-infinite planar dielectric waveguide. As an intermediate step the wavefunction in the exit plane must be calculated. In this case the integral equation technique of Schwinger \cite{14} cannot be implemented, because of the complicated nature of the integral kernel. We evaluate the emitted radiation and the coefficients of reflection back into the waveguide approximately in a first step of an iterative scheme.

We show in two appendices that the iterative scheme converges to the exact solution in the related problems of reflection by a step potential in one-dimensional quantum mechanics and of Fresnel reflection of an electromagnetic plane wave by a half-space. For the planar dielectric waveguide it does not seem practically possible to go beyond the first step of the iterative scheme.

In a numerical example we study a planar waveguide consisting  of a slab of uniform dielectric constant, bounded on both sides by a medium with a smaller dielectric constant \cite{15,16,17}. In the case studied the first step of the iterative scheme leads to a modification of the wavefunction at the exit plane which is relatively small in comparison with the incident wave. Hence we may expect that the calculation provides a reasonable approximation to the exact solution.

\section{\label{II} Planar open end geometry}

We employ Cartesian coordinates $(x,y,z)$ and consider a planar waveguide in the half-space $z<0$ with stratified dielectric constant $\varepsilon(x)$ and uniform magnetic permeability $\mu_1$. In the half-space $z>0$ the dielectric constant is uniform with value $\varepsilon'$ and the magnetic permeability is $\mu_1$.
We consider solutions of Maxwell's equations which do not depend on the coordinate $y$ and depend on time $t$ through a factor $\exp(-i\omega t)$. Waves traveling to the right in the left half-space will be partly reflected at the plane $z=0$, and partly transmitted into the right half-space. The solutions of Maxwell's equations may be decomposed according to two polarizations . For TE-polarization the components $E_x,\;E_z$, and $H_y$ vanish, and the equations may be combined into the single equation
\begin{equation}
\label{2.1}\frac{\partial^2E_y}{\partial x^2}+\frac{\partial^2E_y}{\partial z^2}+\varepsilon\mu_1k^2E_y=0\qquad (\mathrm{TE}).
\end{equation}
We have used gaussian units, and $k=\omega/c$ is the vacuum wavenumber.
For TM-polarization the components $E_y,\;H_x$, and $H_z$ vanish, and the equations may be combined into the single equation
\begin{equation}
\label{2.2}\frac{\partial^2H_y}{\partial x^2}+\frac{\partial^2H_y}{\partial z^2}-\frac{1}{\varepsilon}\frac{d\varepsilon}{dx}\frac{\partial H_y}{\partial x}+\varepsilon\mu_1k^2H_y=0\qquad (\mathrm{TM}).
\end{equation}
We assume that the profile $\varepsilon(x)$ is symmetric, $\varepsilon(-x)=\varepsilon(x)$, and has a simple form with $\varepsilon(x)$ increasing monotonically for $x<0$ from value $\varepsilon_1$ to a maximum value $\varepsilon_2$ at $x=0$. An example of the geometry under consideration is shown in Fig. 1. In the example the dielectric constant in the left half-space equals a constant $\varepsilon_2$ for $-d<x<d$ and a constant $\varepsilon_1<\varepsilon_2$ for $x<-d$ and $x>d$.

For definiteness we consider only TM-polarization. It is convenient to denote the magnetic field component $H_y(x,z)$ for $z<0$ as $u(x,z)$ and for $z>0$ as $v(x,z)$. The continuity conditions at $z=0$ are
\begin{equation}
\label{2.3}u(x,0-)=v(x,0+),\qquad\frac{1}{\varepsilon(x)}\frac{\partial u(x,z)}{\partial z}\bigg|_{z=0-}=\frac{1}{\varepsilon'}\frac{\partial v(x,z)}{\partial z}\bigg|_{z=0+}.
\end{equation}
We consider a solution $u_{0n}(x,z)$ of Eq. (2.2) given by a guided mode solution
\begin{equation}
\label{2.4}u_{0n}(x,z)=\psi_n(x)\exp(ip_nz),
\end{equation}
where $\psi_n(x)$ is the guided mode wavefunction, and $p_n$ the guided mode wavenumber. We assume $p_n>0$, so that the wave $u_{0n}(x,z)\exp(-i\omega t)$ is traveling to the right. The complete solution takes the form
 \begin{equation}
\label{2.5}u_{n}(x,z)=u_{0n}(x,z)+u_{1n}(x,z),\qquad v_n(x,z),
\end{equation}
where $u_{1n}(x,z)$ and $v_n(x,z)$ must be determined such that the continuity conditions Eq. (2.3) are satisfied. The function $u_{1n}(x,z)$ describes the reflected wave, and $v_n(x,z)$ describes the wave radiated into the right-hand half-space.

Since the right-hand half-space is uniform the solution $v_n(x,z)$ takes a simple form, and can be expressed as
 \begin{equation}
\label{2.6}v_n(x,z)=\int^\infty_{-\infty}F_n(q)\exp(iqx+i\sqrt{\varepsilon'\mu_1k^2-q^2}\;z)\;dq.
\end{equation}
The contribution from the interval $-\sqrt{\varepsilon'\mu_1}|k|<q<\sqrt{\varepsilon'\mu_1}|k|$ corresponds to waves traveling to the right, the contribution from $|q|>\sqrt{\varepsilon'\mu_1}|k|$ corresponds to evanescent waves.

Similarly the solution $u_{1n}(x,z)$ in the left half-space can be expressed as
 \begin{equation}
\label{2.7}u_{1n}(x,z)=\sum^{n_m-1}_{m=0}R_{mn}\psi_m(x)\exp(-ip_mz)
+\int^\infty_0R_n(q)\psi(q,x)\exp(-i\sqrt{\varepsilon_1\mu_1k^2-q^2}\;z)\;dq,
\end{equation}
where the sum corresponds to guided waves traveling to the left, with $n_m$ the number of such guided modes possible at the given frequency $\omega$, and the integral corresponds to waves radiating towards the left. We require that the mode solutions are normalized such that \cite{18}
 \begin{eqnarray}
\label{2.8}\int^\infty_{-\infty}\frac{\psi^*_m(x)\psi_n(x)}{\varepsilon(x)}\;dx&=&\delta_{mn},\qquad
\int^\infty_{-\infty}\frac{\psi^*_m(x)\psi(q,x)}{\varepsilon(x)}\;dx=0,\nonumber\\
\int^\infty_{-\infty}\frac{\psi^*(q,x)\psi(q',x)}{\varepsilon(x)}\;dx&=&\delta(q-q').
\end{eqnarray}
The guided mode solutions $\{\psi_m(x)\}$ can be taken to be real. Orthogonality follows from Eq. (2.2). We show in the next section how the functions $u_{1n}(x,z)$ and $v_n(x,z)$ may in principle be evaluated from an iterative scheme. The coefficients $\{R_{mn}\}$ and the amplitude function $R_n(q)$ also follow from the scheme.

\section{\label{III} Iterative scheme}

The iterative scheme is based on successive approximations to the scattering solution. Thus we write the exact solution as infinite sums
\begin{equation}
\label{3.1}u_n(x,z)=\sum^\infty_{j=0}u^{(j)}_n(x,z),\qquad v_n(x,z)=\sum^\infty_{j=0}v^{(j)}_n(x,z),
\end{equation}
with the terms $u^{(j+1)}_n(x,z),\;v^{(j+1)}_n(x,z)$ determined from the previous $u^{(j)}_n(x,z),\;v^{(j)}_n(x,z)$. In zeroth approximation we identify $u^{(0)}_n(x,z)$ with the incident wave,
\begin{equation}
\label{3.2}u^{(0)}_n(x,z)=\psi_n(x)\exp(ip_nz).
\end{equation}
The corresponding $v^{(0)}_n(x,z)$ will be determined by continuity at the exit plane $z=0$. From Eq. (3.2) we have $u^{(0)}_n(x,0-)=\psi_n(x)$. This has the Fourier transform
\begin{equation}
\label{3.3}\phi_n(q)=\frac{1}{2\pi}\int^\infty_{-\infty}\psi_n(x)\exp(-iqx)\;dx.
\end{equation}
Using continuity of the wavefunction at $z=0$ and the expression (2.6) we find correspondingly
\begin{equation}
\label{3.4}v^{(0)}_n(x,z)=\int^\infty_{-\infty}\phi_n(q)\exp(iqx+i\sqrt{\varepsilon'\mu_1k^2-q^2}\;z)\;dq,
\end{equation}
so that in zeroth approximation $F^{(0)}_n(q)=\phi_n(q)$. Clearly the zeroth approximation does not satisfy the second continuity equation in Eq. (2.3), and we must take care of this in the next approximation.

For the difference of terms in Eq. (2.3) we find
\begin{equation}
\label{3.5}\rho^{(0)}_n(x)=\frac{-i}{\varepsilon(x)}p_n\psi_n(x)+\frac{i}{\varepsilon'}\int^\infty_{-\infty}\sqrt{\varepsilon'\mu_1k^2-q^2}\;\phi_n(q)\exp(iqx)\;dq.
\end{equation}
By symmetry $\rho^{(0)}_n(x)$ is symmetric in $x$ for $n$ even, antisymmetric in $x$ for $n$ odd.

The next approximation $u^{(1)}_n(x,z)$ can be found by comparison with the solution of the problem where the profile $\varepsilon(x)$ extends over all space and radiation is generated by a source $\varepsilon(x)\rho(x)\delta(z)$ with a Sommerfeld radiation condition, so that radiation is emitted to the right for $z>0$ and to the left for $z<0$. This antenna solution can be expressed as
\begin{equation}
\label{3.6}u_A(x,z)=\int^\infty_{-\infty} K(x,x',z)\rho(x')\;dx',
\end{equation}
with kernel $K(x,x',z)$. The latter can be calculated from the Fourier decomposition
\begin{equation}
\label{3.7}\delta(z)=\frac{1}{2\pi}\int^\infty_{-\infty}e^{ipz}\;dp,
\end{equation}
in terms of the integral
\begin{equation}
\label{3.8}K(x,x',z)=\frac{1}{2\pi}\int^\infty_{-\infty}G(x,x',p)\;e^{ipz}\;dp,
\end{equation}
with the prescription that the path of integration in the complex $p$ plane runs just above the negative real axis and just below the positive real axis. The Green function $G(x,x',p)$ can be found from the solution of the one-dimensional wave equation,
\begin{equation}
\label{3.9}\frac{d^2G}{dx^2}-\frac{1}{\varepsilon}\frac{d\varepsilon}{dx}\frac{dG}{dx}+(\varepsilon\mu_1k^2-p^2)G=\varepsilon(x)\delta(x-x').
\end{equation}
The solution takes the form \cite{18}
\begin{equation}
\label{3.10}G(x,x',p)=\sqrt{\varepsilon(x)}\frac{f_2(x_<,p)f_3(x_>,p)}{\Delta(f_2,f_3,p)}\sqrt{\varepsilon(x')},
\end{equation}
where $x_<(x_>)$ is the smaller (larger) of $x$ and $x'$, and the remaining quantities will be specified in the next section. The Green function satisfies the symmetry properties
\begin{equation}
\label{3.11}G(x,x',-p)=G(x,x',p),\qquad G(-x,-x',p)=G(x,x',p),
\end{equation}
and the reciprocity relation
\begin{equation}
\label{3.12}G(x,x',p)=G(x',x,p).
\end{equation}
Consequently the kernel $K(x,x',z)$ has the properties
\begin{equation}
\label{3.13} K(x,x',-z)=K(x,x',z),\qquad K(-x,-x',z)=K(x,x',z),
\end{equation}
as well as
\begin{equation}
\label{3.14}K(x,x',z)=K(x',x,z).
\end{equation}

The function $u^{(1)}_n(x,z)$ is now identified as
\begin{equation}
\label{3.15}u^{(1)}_n(x,z)=-\int^\infty_{-\infty} K(x,x',z)\rho^{(0)}_n(x')\;dx'.
\end{equation}
The minus sign is needed to provide near cancellation of the source density between the zeroth and first order solutions, $\rho^{(0)}_n(x)+\rho^{(1)}_n(x)\approx 0$.
We find the first order function $F_n^{(1)}(q)$ by Fourier transform from the value at $z=0$ in the form
\begin{equation}
\label{3.16}F^{(1)}_n(q)=\frac{1}{2\pi}\int^\infty_{-\infty}u^{(1)}_n(x,0)e^{-iqx}\;dx.
\end{equation}
The corresponding function $v^{(1)}_n(x,z)$ is found from Eq (2.6). The first order source density $\rho^{(1)}_n(x)$ is found to be
\begin{equation}
\label{3.17}\rho^{(1)}_n(x)=\frac{-1}{\varepsilon(x)}\frac{\partial u^{(1)}_n(x,z)}{\partial z}\bigg|_{z=0}+\frac{i}{\varepsilon'}\int^\infty_{-\infty}\sqrt{\varepsilon'\mu_1k^2-q^2}F^{(1)}_n(q)\exp(iqx)\;dq.
\end{equation}

In principle the first order function $u^{(1)}_n(x,0)$ in the exit plane $z=0$ may be regarded as the result of a linear operator $\mathcal{R}^{(1)}$ acting on the state $\psi_n(x)$ given by the incident wave. The iterated solution then corresponds to the action with the operator $\mathcal{R}=\mathcal{R}^{(1)}(\mathcal{I}-\mathcal{R}^{(1)})^{-1}$, where $\mathcal{I}$ is the identity operator. In order $j$ of the geometric series corresponding to the operator $\mathcal{R}$ the wavefunctions $u^{(j)}_{1n}(x,z)$ and $v^{(j)}_n(x,z)$ in the left and right half-space can be found by completing the function $u^{(j)}_{1n}(x,0)=v^{(j)}_{1n}(x,0)$ in the exit plane by left and right running waves respectively.

Assuming that the scheme has been extended to all orders we obtain the solutions $u_{1n}(x,z)=u_n(x,z)-u^{(0)}_n(x,z)$ and $v_n(x,z)$ given by Eq. (3.1). By construction at each step $u^{(j)}_n(x,0)=v^{(j)}_n(x,0)$. In the limit we must have
\begin{equation}
\label{3.18}\sum^\infty_{j=0}\rho^{(j)}_n(x)=0,
\end{equation}
so that the continuity conditions Eq. (2.3) are exactly satisfied. In Appendix A we show how the iterative scheme reproduces the exact solution for reflection from a step potential in one-dimensional quantum mechanics. In Appendix B we show the same for Fresnel reflection.

In the integral in Eq. (3.15) it is convenient to perform the integral over $p$ first, since $\rho^{(0)}_n(x')$ does not depend on $p$. The pole at $-p_m$, arising from a zero of the denominator $\Delta$ in Eq. (3.10), yields the first order reflection coefficient \cite{18}
\begin{equation}
\label{3.19}R^{(1)}_{mn}=\frac{i}{2p_m}\int^\infty_{-\infty}\psi_m(x)\rho^{(0)}_n(x)\;dx.
\end{equation}
The second term in Eq. (2.7) corresponds to the remainder of the integral, after subtraction of the simple pole contributions. The function $R^{(1)}_n(q)$ will be discussed in the next section.
In the calculation of $F_n^{(1)}(q)$ from Eq. (3.16) we find
\begin{equation}
\label{3.20}F_n^{(1)}(q)=\sum^{n_m-1}_{m=0}R^{(1)}_{mn}\phi_m(q)+\delta F_n^{(1)}(q),
\end{equation}
where $\delta F_n^{(1)}(q)$ is the contribution from the remainder of the integral over $p$, after subtraction of the simple pole contributions.

Formally, in the complete solution Eq. (2.7) the reflection coefficients $R_{mn}$ and the amplitude function $R_n(q)$ are found as
 \begin{equation}
\label{3.21}R_{mn}=(\psi_m,u_{1n}(0)),\qquad R_n(q)=(\psi(q),u_{1n}(0)).
\end{equation}
with the scalar product as given by Eq. (2.8). The first step of the iterative scheme yields
 \begin{equation}
\label{3.22}R^{(1)}_{mn}=(\psi_m,u^{(1)}_n(0)),\qquad R^{(1)}_n(q)=(\psi(q),u^{(1)}_n(0)).
\end{equation}
The continuum states $\psi(q)$ can be discretized in the usual way, so that the expressions in Eq. (3.22) can be regarded as elements of a matrix $\du{R}^{(1)}$. As indicated above, the iterative scheme corresponds to a geometric series, so that the reflection coefficients in Eq. (3.21) can be found as elements of the matrix
 \begin{equation}
\label{3.23}\du{R}=(\du{I}-\du{R}^{(1)})^{-1}-\du{I},
\end{equation}
where $\du{I}$ is the identity matrix.

Finally the function $F_n(q)$ can be found from the corresponding state $u_n(0)$  as in Eq. (3.16). The function $F_n(q)$ can be related to the radiation scattered into the right half-space. The scattering angle $\theta$ is related to the component $q$ by
 \begin{equation}
\label{3.24}\sin\theta=q/\sqrt{\varepsilon'\mu_1k^2}.
\end{equation}
Defining the scattering cross section $\sigma_n(\theta)$ by
 \begin{equation}
\label{3.25}\sigma_n(\theta)\sin\theta d\theta=|F_n(q)|^2qdq,
\end{equation}
we find the relation
 \begin{equation}
\label{3.26}\sigma_n(\theta)=\sqrt{\varepsilon'\mu_1k^2}\sqrt{\varepsilon'\mu_1k^2-q^2}\;|F_n(q)|^2.
\end{equation}
In lowest approximation the cross section is proportional to the absolute square of the Fourier transform of the guided mode $\psi_n(x)$. To higher order the cross section is affected by the reflection into other modes.

\section{\label{IV} Continuous spectrum}

The calculation of the function $R^{(1)}_n(q)$ corresponding to the contribution from the continuous spectrum requires a separate discussion. The wave equation (3.9) is related to a quantummechanical Schr\"odinger equation for a particle in a potential. The bound states of the Schr\"odinger problem correspond to the guided modes, and the scattering states correspond to a continuous spectrum of radiation modes. The eigenstates of the Hamilton operator of the Schr\"odinger problem satisfy a completeness relation which can be usefully employed in the waveguide problem.

Explicitly the homogeneous one-dimensional Schr\"odinger equation corresponding to Eq. (3.9) via the relation $\psi(x)=\sqrt{\varepsilon(x)}f(x)$ reads \cite{18}
  \begin{equation}
\label{4.1}\frac{d^2f}{dx^2}-V(x)f=p^2f,
\end{equation}
where the function $V(x)$ is given by
  \begin{equation}
\label{4.2}V(x)=-\varepsilon\mu_1k^2+\sqrt{\varepsilon}\frac{d^2}{dx^2}\frac{1}{\sqrt{\varepsilon}}.
\end{equation}
By comparison with the quantummechanical Schr\"odinger equation we see that
  \begin{equation}
\label{4.3}U(x)=k_1^2+V(x),
\end{equation}
where $k_1=\sqrt{\varepsilon_1\mu_1}\;k$, may be identified as the potential. The bound state energies correspond to $\{k_1^2-p_n^2\}$.

It is convenient to assume that the dielectric profile $\varepsilon(x)$ equals $\varepsilon_1$ for $x<-x_1$ and $x>x_1$, so that the potential $U(x)$ vanishes for $|x|>x_1$.
We define three independent solutions of the Schr\"odinger equation (4.1) with specified behavior for $|x|>x_1$. The behavior of the function $f_1(x,p)$ is specified as
\begin{eqnarray}
\label{4.4}f_1(x,p)&=&e^{iq_1x},\qquad\mathrm{for}\;x<-x_1,\nonumber\\
f_1(x,p)&=&W_{11}e^{iq_1x}+W_{21}e^{-iq_1x},\qquad\mathrm{for}\;x>x_1,
\end{eqnarray}
with wavenumber
\begin{equation}
\label{4.5}q_1=\sqrt{k_1^2-p^2}.
\end{equation}
The behavior of the function $f_2(x,p)$ is specified as
\begin{eqnarray}
\label{4.6}f_2(x,p)&=&e^{-iq_1x},\qquad\mathrm{for}\;x<-x_1,\nonumber\\
f_2(x,p)&=&W_{12}e^{iq_1x}+W_{22}e^{-iq_1x},\qquad\mathrm{for}\;x>x_1.
\end{eqnarray}
Similarly, the behavior of the function $f_3(x,p)$ is specified as
\begin{eqnarray}
\label{4.7}f_3(x,p)&=&W_{22}e^{iq_1x}+W_{12}e^{-iq_1x},\qquad\mathrm{for}\;x<-x_1,\nonumber\\
f_3(x,p)&=&e^{iq_1x},\qquad\mathrm{for}\;x>x_1.
\end{eqnarray}
The coefficients $W_{12}$ and $W_{22}$ are elements of the transfer matrix of the planar structure,
\begin{equation}
\label{4.8}\du{W}=\left(\begin{array}{cc}W_{11}&W_{12}\\
W_{21}&W_{22}\end{array}\right)=\frac{1}{T'}\left(\begin{array}{cc}TT'-RR'&R'\\
-R&1\end{array}\right).
\end{equation}
Because of the assumed symmetry of the dielectric profile we have in the present case $R'=R$ and $T'=T$, so that $W_{21}=-W_{12}$. Moreover
\begin{equation}
\label{4.9}W_{11}W_{22}+W_{12}^2=1.
\end{equation}
The functions $f_2(x,p)$ and $f_3(x,p)$ were used in the calculation of the Green function in Eq. (3.10). The denominator in that expression is given by
\begin{equation}
\label{4.10}\Delta(f_2,f_3)=2iq_1W_{22}(p,k).
\end{equation}
From the solution of the inhomogeneous Schr\"odinger equation it follows that the completeness relation of the normal mode solutions may be expressed as \cite{19}
\begin{equation}
\label{4.11}\sum^{n_m-1}_{n=0}\frac{\psi_n(x)\psi_n(x')}{\sqrt{\varepsilon(x)\varepsilon(x')}}
+\frac{1}{2\pi}\int^\infty_{-\infty}\frac{f_2(x,\sqrt{k_1^2-q^2})f^*_2(x',\sqrt{k_1^2-q^2})}{|W_{22}(\sqrt{k_1^2-q^2},k)|^2}\;dq=\delta(x-x').
\end{equation}
Correspondingly the Green function may be decomposed as
\begin{eqnarray}
\label{4.12}G(x,x',p)&=&\sum^{n_m-1}_{n=0}\frac{\psi_n(x)\psi_n(x')}{p_n^2-p^2}\nonumber\\
&+&\frac{1}{2\pi}\sqrt{\varepsilon(x)\varepsilon(x')}\int^\infty_{-\infty}\frac{f_2(x,\sqrt{k_1^2-q^2})f^*_2(x',\sqrt{k_1^2-q^2})}{(k_1^2-q^2-p^2)\;|W_{22}(\sqrt{k_1^2-q^2},k)|^2}\;dq.
\end{eqnarray}
Hence we find for the antenna kernel $K(x,x',z)$ from Eq. (3.8) by use of the integration prescription
\begin{eqnarray}
\label{4.13}K(x,x',z)&=&\sum^{n_m-1}_{n=0}\frac{-i}{2p_n}\;\psi_n(x)\psi_n(x')\;e^{ip_n|z|}\nonumber\\
&-&\frac{i}{4\pi}\sqrt{\varepsilon(x)\varepsilon(x')}
\int^\infty_0\frac{f_2(x,\sqrt{k_1^2-q^2})f^*_2(x',\sqrt{k_1^2-q^2})}{\sqrt{k_1^2-q^2}\;|W_{22}(\sqrt{k_1^2-q^2},k)|^2}\;e^{i\sqrt{k_1^2-q^2}|z|}\;dq.\nonumber\\
\end{eqnarray}
The first order left-hand wavefunction is therefore found from Eq. (3.15) as
\begin{eqnarray}
\label{4.14}u^{(1)}_n(x,z)&=&\sum^{n_m-1}_{m=0}\frac{i}{2p_m}(\psi_m,\varepsilon \rho^{(0)}_n)\psi_m(x)\;e^{-ip_mz}\nonumber\\
&+&\frac{i}{4\pi}\sqrt{\varepsilon(x)\varepsilon(x')}
\int^\infty_0\frac{(\sqrt{\varepsilon}f_2(\sqrt{k_1^2-q^2}),\varepsilon \rho^{(0)}_n)}{\sqrt{k_1^2-q^2}\;|W_{22}(\sqrt{k_1^2-q^2},k)|^2}\;f_2(x,\sqrt{k_1^2-q^2})e^{-i\sqrt{k_1^2-q^2}z}\;dq,\nonumber\\
\end{eqnarray}
with scalar product as given by Eq. (2.8). The wavefunction is the sum of guided modes running to the left and of radiation into the left-hand half-space. The first term agrees with the reflection coefficient given by Eq. (3.19). By symmetry the matrix element $(\psi_m,\varepsilon \rho^{(0)}_n)$ vanishes unless $m$ and $n$ are both even or both odd. The integral provides an alternative expression for the remainder $\delta u^{(1)}_n(x,z)$.

From Eq. (4.11) we may identify
\begin{equation}
\label{4.15}\psi(q,x)=\frac{1}{\sqrt{2\pi}|W_{22}(\sqrt{k_1^2-q^2},k)|}\;\sqrt{\varepsilon(x)}f_2(x,\sqrt{k_1^2-q^2}).
\end{equation}
With this definition the function $R^{(1)}_n(q)$ is given by
\begin{eqnarray}
\label{4.16}R^{(1)}_n(q)&=&\frac{i}{2\sqrt{2\pi}}\;\frac{(\sqrt{\varepsilon}f_2(\sqrt{k_1^2-q^2}),\varepsilon \rho^{(0)}_n)}{\sqrt{k_1^2-q^2}|W_{22}(\sqrt{k_1^2-q^2},k)|}\nonumber\\
&=&\frac{i}{2\sqrt{k^2_1-q^2}}(\psi(q),\varepsilon \rho^{(0)}_n).
\end{eqnarray}
The second line has the same structure as Eq. (3.19). Although the decomposition in Eq. (4.14) is of theoretical interest, the calculation of the function $u^{(1)}_n(x,z)$ is performed more conveniently as indicated in Eq. (3.15), with the integral over $p$ performed first.

\section{\label{V} Numerical example}

We demonstrate the effectiveness of the scheme on a numerical example. We consider a flat dielectric profile defined by $\varepsilon(x)=\varepsilon_2$ for $-d<x<d$ and $\varepsilon(x)=\varepsilon_1$ for $|x|>d$ with values $\varepsilon_2=2.25$ and $\varepsilon_1=2.13$. In the right half-space we put $\varepsilon'=1$, and we put $\mu_1=1$  everywhere. The geometry is shown in Fig. 1.

By symmetry the guided modes in infinite space are either symmetric or antisymmetric in $x$. The explicit expressions for the mode wavefunctions can be found by the transfer matrix method \cite{18}. At each of the two discontinuities the coefficients of the plane waves $\exp(iq_ix)$ and $\exp(-iq_ix)$ are transformed into coefficients of the plane waves $\exp(iq_jx)$ and $\exp(-iq_jx)$ by a matrix involving Fresnel coefficients given by
 \begin{equation}
\label{5.1}t_{ij}=\frac{2\varepsilon_jq_i}{\varepsilon_iq_j+\varepsilon_jq_i},\qquad r_{ij}=\frac{\varepsilon_jq_i-\varepsilon_iq_j}{\varepsilon_iq_j+\varepsilon_jq_i},\qquad (i,j)=(1,2),
\end{equation}
with wavenumbers
 \begin{equation}
\label{5.2}q_j=\sqrt{k^2_j-p^2},\qquad k_j=\sqrt{\varepsilon_j\mu_1}\;k.
\end{equation}
The wavenumbers $p_n(k)$ of the guided modes are found as zeros of the transfer matrix element $W_{22}(p,k)$, which takes the explicit form
 \begin{equation}
\label{5.3}W_{22}(p,k)=e^{2iq_1d}\bigg[\cos2q_2d-i\;\frac{\varepsilon^2_1q^2_2+\varepsilon^2_2q^2_1}{2\varepsilon_1\varepsilon_2q_1q_2}\;\sin2q_2d\bigg].
\end{equation}
The guided mode wavefunctions $\{\psi_n(x)\}$, their Fourier transforms $\{\phi_n(q)\}$, and the Green function $G(x,x',p)$ can be found in explicit form.
In Fig. 2 we show the ratio of wavenumbers $p_n(k)/k$ as a function of $kd$ for the first few guided modes.

We choose the frequency corresponding to $kd=12$. In that case there are two symmetric modes, denoted as TM0 and TM2, and one antisymmetric mode, denoted as TM1. We assume the incident wave to be symmetric in $x$. Then it is not necessary to consider the antisymmetric mode. In Fig. 3 we show the corresponding normalized wavefunctions $\psi_0(x)$ and $\psi_2(x)$. In Fig. 4 we show their Fourier transforms $\phi_0(q)$ and $\phi_2(q)$. The wavenumbers at $kd=12$ are $p_0=17.955/d$ and $p_2=17.624/d$. The edge of the continuum is given by $k_1d=17.513$, and the corresponding value for $\varepsilon_2$ is $k_2d=18$.

In Fig. 5 we show the source density $-i\rho^{(0)}_n(x)$ of the zeroth approximation for $n=0,\;2$ as a function of $x$, as given by Eq. (3.5). The coefficients of the simple pole contributions can be calculated from Eq. (3.19). We find numerically for the discrete part $\du{R}^{(1)}_d$ of the matrix $\du{R}^{(1)}$
\begin{equation}
\label{5.4}\du{R}^{(1)}_d=\left(\begin{array}{cc}R^{(1)}_{00}&R^{(1)}_{02}\\
R^{(1)}_{20}&R^{(1)}_{22}\end{array}\right)=\left(\begin{array}{cc}-0.2477&-0.0004\\
0.0013&-0.2312\end{array}\right).
\end{equation}

For the first correction to the emitted radiation we need to calculate the function $u^{(1)}_n(x,0)$. The kernel $K(x,x',0)$ in Eq. (3.15) can be evaluated numerically. On account of the symmetry in $\pm p$ it is sufficient to calculate twice the integral along the positive real $p$ axis, with path of integration just below the axis. In the numerical integration over $p$ in Eq. (3.8) the simple poles at $\{p_m\}$ cause problems. In order to avoid the simple poles we therefore integrate instead along a contour consisting of the line from $0$ to $k_1$ just below the axis, a semi-circle in the lower half of the complex $p$ plane centered at $(k_1+k_2)/2$ and of radius $(k_2-k_1)/2$, and the line just below the real axis from $k_2$ to $+\infty$. In Fig. 6 we plot as an example the real part of $K(x,0,0)$ as a function of $x$. The plot of the imaginary part is similar.

In Fig. 7 we show the imaginary part of the function $u^{(1)}_0(x,0)$, as calculated from Eq. (3.15). This is nearly identical with the contribution from the simple poles at $p_0$ and $p_2$, which is also shown in Fig. 7. In Fig. 8 we show the real part of the function $u^{(1)}_0(x,0)$. Here the simple poles do not contribute. The magnitude of the wavefunction at the origin $u^{(1)}_0(0,0)=-0.149-0.005i$ may be compared with that of the zeroth approximation $u^{(0)}_0(0,0)=1.346$. This shows that the first order correction is an order of magnitude smaller than the zeroth order approximation. Consequently we may expect that the sum $u^{(0)}_0(x,0)+u^{(1)}_0(x,0)$ provides a close approximation to the exact value.

In Fig. 9 we show the absolute value $|F^{(0)}_0(q)+F^{(1)}_0(q)|$ of the Fourier transform of the sum $u^{(0)}_0(x,0)+u^{(1)}_0(x,0)$, and compare with the zeroth approximation $|F^{(0)}_0(q)|=|\phi_0(q)|$. By use of Eq. (3.26) the absolute square of the transform yields the angular distribution of radiation emitted into the right-hand half-space.

\section{\label{VI} Discussion}

In the above we have employed an iterative scheme inspired by the exact solution of two fundamental scattering problems, reflection by a step potential in one-dimensional quantum mechanics, shown in Appendix A, and Fresnel reflection of electromagnetic radiation by a half-space, shown in Appendix B. For the planar dielectric waveguide we have implemented only the first step of the iterative scheme. In the numerical example shown in Sec. V even this first step leads to interesting results. The method is sufficiently successful that it encourages application in other situations.

In particular it will be of interest to apply the method to a circular cylindrical dielectric waveguide or optical fiber. The mathematics of the method carries over straightforwardly to this more complicated geometry, with the plane wave behavior in the transverse direction replaced by Bessel functions.

Due to symmetry the problem for both planar and cylindrical geometry can be reduced to an equation for a scalar wavefunction, so that the theory is similar to that for sound propagation. This suggests that an interesting comparison can be made with a lattice Boltzmann simulation. For a rigid circular pipe such a simulation has already been performed by da Silva and Scavone \cite{20}, with interesting results. A finite element method has been applied to a rigid open-ended duct of more general cross section \cite{21}.

\newpage
\appendix
\section{}

In this Appendix we show how the iterative scheme reproduces the exact solution of the time-independent one-dimensional Schr\"odinger equation with a step potential. We consider the equation
 \begin{equation}
\label{A.1}-\frac{d^2u}{dz^2}+V(z)u=p^2u,
\end{equation}
with potential $V(z)=0$ for $z<0$ and $V(z)=V$ for $z>0$. In proper units $p^2$ is the energy. We denote the solution for $z>0$ as $v(z)$. For a wave incident from the left the exact solution reads
 \begin{equation}
\label{A.2}u(z)=e^{ipz}+Be^{-ipz},\qquad v(z)=Ce^{ip'z},
\end{equation}
where $p'=\sqrt{p^2-V}$, and the reflection coefficient $B$ and transmission coefficient $C$ are given by
 \begin{equation}
\label{A.3}B=\frac{p-p'}{p+p'},\qquad C=\frac{2p}{p+p'}.
\end{equation}
The wavefunction and its derivative are continuous at $z=0$.

We apply the iterative scheme and put to zeroth order
 \begin{equation}
\label{A.4}u^{(0)}(z)=e^{ipz},\qquad v^{(0)}(z)=e^{ip'z}.
\end{equation}
The antenna solution $u_A(z)$ solves the equation
 \begin{equation}
\label{A.5}\frac{d^2u_A}{dz^2}+p^2u_A=\rho\delta(z)
\end{equation}
for all $z$. It is given by
 \begin{equation}
\label{A.6}u_A(z)=K(z)\rho,\qquad K(z)=\frac{1}{2ip}\;e^{ip|z|}.
\end{equation}
To zeroth order the source $\rho$ is
 \begin{equation}
\label{A.7}\rho^{(0)}=-\frac{du^{(0)}}{dz}\bigg|_{z=0}+\frac{dv^{(0)}}{dz}\bigg|_{z=0}=-i(p-p').
\end{equation}
We put the first order solution equal to
 \begin{equation}
\label{A.8}u^{(1)}(z)=-K(z)\rho^{(0)}=\frac{p-p'}{2p}\;e^{-ipz},\qquad v^{(1)}(z)=\frac{p-p'}{2p}\;e^{ip'z}.
\end{equation}
Note the minus sign in $-K(z)\rho^{(0)}$. The value at the exit $z=0$ is sufficient to calculate the coefficients $B$ and $C$ from the geometric series
 \begin{equation}
\label{A.9}B=\sum^\infty_{j=1}\bigg(\frac{p-p'}{2p}\bigg)^j,\qquad C=\sum^\infty_{j=0}\bigg(\frac{p-p'}{2p}\bigg)^j.
\end{equation}
By continuation one finds for the wave function at order $j$ for $j\geq 1$
 \begin{equation}
\label{A.10}u^{(j)}(z)=\bigg(\frac{p-p'}{2p}\bigg)^{j}e^{-ipz},\qquad v^{(j)}(z)=\bigg(\frac{p-p'}{2p}\bigg)^{j}e^{ip'z}.
\end{equation}
Hence for $j\geq 1$ the source at order $j$ is
 \begin{equation}
\label{A.11}\rho^{(j)}=i(p+p')\bigg(\frac{p-p'}{2p}\bigg)^{j},\qquad (j\geq 1)
\end{equation}
 so that the sum over all $j$ vanishes,
\begin{equation}
\label{A.12}\sum^\infty_{j=0}\rho^{(j)}=0,
\end{equation}
as it should. Alternatively one can write directly from Eq. (A8)
 \begin{equation}
\label{A.13}u_1(z)=\frac{(p-p')/(2p)}{1-(p-p')/(2p)}\;e^{-ipz}=\frac{p-p'}{p+p'}\;e^{-ipz},\qquad v_1(z)=\frac{p-p'}{p+p'}\;e^{ip'z}.
\end{equation}
Adding this to $u^{(0)}(z),\;v^{(0)}(z)$ one reproduces Eq. (A2).

\section{}

In this Appendix we show how the iterative scheme reproduces the exact solution for Fresnel reflection from a half-space. We consider infinite space with dielectric constant $\varepsilon$ for $z<0$ and $\varepsilon'$ for $z>0$. The magnetic permeability equals $\mu_1$ everywhere. We consider waves independent of $y$ and TM-polarization. Then the magnetic field component $H_y(x,z)$ satisfies the scalar equation Eq. (2.2). We put $H_y(x,z)=u(x,z)$ for $z<0$ and $H_y(x,z)=v(x,z)$ for $z>0$. The continuity conditions at $z=0$ are
 \begin{equation}
\label{B.1}u(x,0-)=v(x,0+),\qquad\frac{1}{\varepsilon}\frac{\partial u(x,z)}{\partial z}\bigg|_{z=0-}=\frac{1}{\varepsilon'}\frac{\partial v(x,z)}{\partial z}\bigg|_{z=0+}.
\end{equation}
 For a plane wave incident from the left the exact solution reads
 \begin{equation}
\label{B.2}u(x,z)=e^{iqx}\big[e^{ipz}+Be^{-ipz}\big],\qquad v(x,z)=Ce^{iqx+ip'z},
\end{equation}
with $p=\sqrt{\varepsilon\mu_1k^2-q^2},\;p'=\sqrt{\varepsilon'\mu_1k^2-q^2}$, and reflection coeficient $B$ and transmission coefficient $C$ given by
 \begin{equation}
\label{B.3}B=\frac{\varepsilon' p-\varepsilon p'}{\varepsilon' p+\varepsilon p'},\qquad C=\frac{2\varepsilon' p}{\varepsilon' p+\varepsilon p'}.
\end{equation}

We apply the iterative scheme and put to zeroth order
 \begin{equation}
\label{B.4}u^{(0)}(x,z)=e^{iqx+ipz},\qquad v^{(0)}(x,z)=e^{iqx+ip'z}.
\end{equation}
The antenna solution $u_A(x,z)$ solves the equation
 \begin{equation}
\label{B.5}\frac{\partial^2u_A}{\partial x^2}+\frac{\partial^2u_A}{\partial z^2}+\varepsilon\mu_1k^2u_A=\varepsilon \rho(x)\delta(z)
\end{equation}
for all $(x,z)$. For $\rho(x)=\rho_qe^{iqx}$ it is given by
 \begin{equation}
\label{B.6}u_A(x,z)=e^{iqx}K(z)\rho_q,\qquad K(z)=\frac{\varepsilon}{2ip}\;e^{ip|z|}.
\end{equation}
To zeroth order the source $\rho_q$ is
 \begin{equation}
\label{B.7}\rho_q^{(0)}=e^{-iqx}\bigg[-\frac{1}{\varepsilon}\frac{\partial u^{(0)}}{\partial z}\bigg|_{z=0}+\frac{1}{\varepsilon'}\frac{\partial v^{(0)}}{\partial z}\bigg|_{z=0}\bigg]=-i\bigg(\frac{p}{\varepsilon}-\frac{p'}{\varepsilon'}\bigg).
\end{equation}
We put the first order solution equal to
 \begin{eqnarray}
\label{B.8}u^{(1)}(x,z)&=&-e^{iqx}K(z)\rho^{(0)}_q=\frac{\varepsilon'p-\varepsilon p'}{2\varepsilon'p}\;e^{iqx-ipz},\nonumber\\
v^{(1)}(x,z)&=&\frac{\varepsilon'p-\varepsilon p'}{2\varepsilon'p}\;e^{iqx+ip'z}.
\end{eqnarray}
Note the minus sign in $-e^{iqx}K(z)\rho^{(0)}_q$. The value at the exit $z=0$ is sufficient to calculate the coefficients $B$ and $C$ from the geometric series
 \begin{equation}
\label{B.9}B=\sum^\infty_{j=1}\bigg(\frac{\varepsilon'p-\varepsilon p'}{2\varepsilon'p}\bigg)^j,\qquad C=\sum^\infty_{j=0}\bigg(\frac{\varepsilon'p-\varepsilon p'}{2\varepsilon'p}\bigg)^j.
\end{equation}
By continuation one finds for the wave function at order $j$ for $j\geq 1$
 \begin{equation}
\label{B10}u^{(j)}(x,z)=\bigg(\frac{\varepsilon'p-\varepsilon p'}{2\varepsilon'p}\bigg)^{j}e^{iqx-ipz},\qquad v^{(j)}(x,z)=\bigg(\frac{\varepsilon'p-\varepsilon p'}{2\varepsilon'p}\bigg)^{j}e^{iqx+ip'z}.
\end{equation}
Hence for $j\geq 1$ the source at order $j$ is
 \begin{equation}
\label{B11}\rho^{(j)}_q=i\bigg(\frac{p}{\varepsilon}+\frac{p'}{\varepsilon'}\bigg)\bigg(\frac{\varepsilon'p-\varepsilon p'}{2\varepsilon'p}\bigg)^{j},\qquad (j\geq 1)
\end{equation}
 so that the sum over all $j$ vanishes,
\begin{equation}
\label{B12}\sum^\infty_{j=0}\rho^{(j)}_q=0,
\end{equation}
as it should. Alternatively one can write directly from Eq. (B8)
 \begin{eqnarray}
\label{B13}u_1(x,z)&=&\frac{(\varepsilon'p-\varepsilon p')/(2\varepsilon'p)}{1-(\varepsilon'p-\varepsilon p')/(2\varepsilon'p)}\;e^{iqx-ipz}=\frac{\varepsilon'p-\varepsilon p'}{\varepsilon'p+\varepsilon p'}\;e^{iqx-ipz},\nonumber\\
v_1(x,z)&=&\frac{\varepsilon'p-\varepsilon p'}{\varepsilon'p+\varepsilon p'}\;e^{iqx+ip'z}.
\end{eqnarray}
Adding this to $u^{(0)}(x,z),\;v^{(0)}(x,z)$ one reproduces Eq. (B2).

We note that the zeroth and first order source densities are related by
\begin{equation}
\label{B14}\rho^{(1)}_q=-M\rho^{(0)}_q,\qquad M=\frac{\varepsilon'p+\varepsilon p'}{2\varepsilon'p}.
\end{equation}
Hence we find
\begin{equation}
\label{B15}B=1-M^{-1},\qquad C=1+B.
\end{equation}
This suggests that more generally the complete solution of the scattering problem may be found from the relation between the zeroth and first order source densities.

\newpage

\section*{Figure captions}

\subsection*{Fig. 1}
Geometry of the planar waveguide.

\subsection*{Fig. 2}
Plot of the reduced wavenumber $p_n(k)/k$ of the lowest order guided waves for $n=0,1,2$, as functions of $kd$ for values of the dielectric constant given in the text.

\subsection*{Fig. 3}
Plot of the wavefunctions $\psi_0(x)$ and $\psi_2(x)$ of the guided modes with $n=0$ (no nodes) and $n=2$ (two nodes)  as functions of $x/d$.

\subsection*{Fig. 4}
Plot of the Fourier transform $\phi_0(q)$ and $\phi_2(q)$ of the wavefunctions of the guided modes with $n=0$ (solid curve) and $n=2$ (dashed curve)  as functions of $qd$.

\subsection*{Fig. 5}
Plot of the source densities $-i\rho^{(0)}_0(x)$ and $-i\rho^{(0)}_2(x)$, as given by Eq. (3.5), as functions of $x/d$.

\subsection*{Fig. 6}
Plot of the real part of the kernel $K(x,0,0)$, as given by Eq. (3.8), as a function of $x/d$.

\subsection*{Fig. 7}
Plot of the real part of the first order wavefunction $u^{(1)}_0(x,0)$ at the exit plane as a function of $x/d$ (solid curve), compared with the contribution of the two guided waves $R^{(1)}_{00}u_0(x)+R^{(1)}_{20}u_2(x)$ (dashed curve).

\subsection*{Fig. 8}
Plot of the imaginary part of the first order wavefunction $u^{(1)}_0(x,0)$ at the exit plane as a function of $x/d$.

\subsection*{Fig. 9}
Plot of the absolute value of the Fourier transform $|F^{(0)}_0(q)+F^{(1)}_0(q)|$ of the sum of zero order and first order wave function at the exit plane (solid curve), compared with the Fourier transform $|F^{(0)}_0(q)|$ (dashed curve).
\newpage
\setlength{\unitlength}{1cm}
\begin{figure}
 \includegraphics{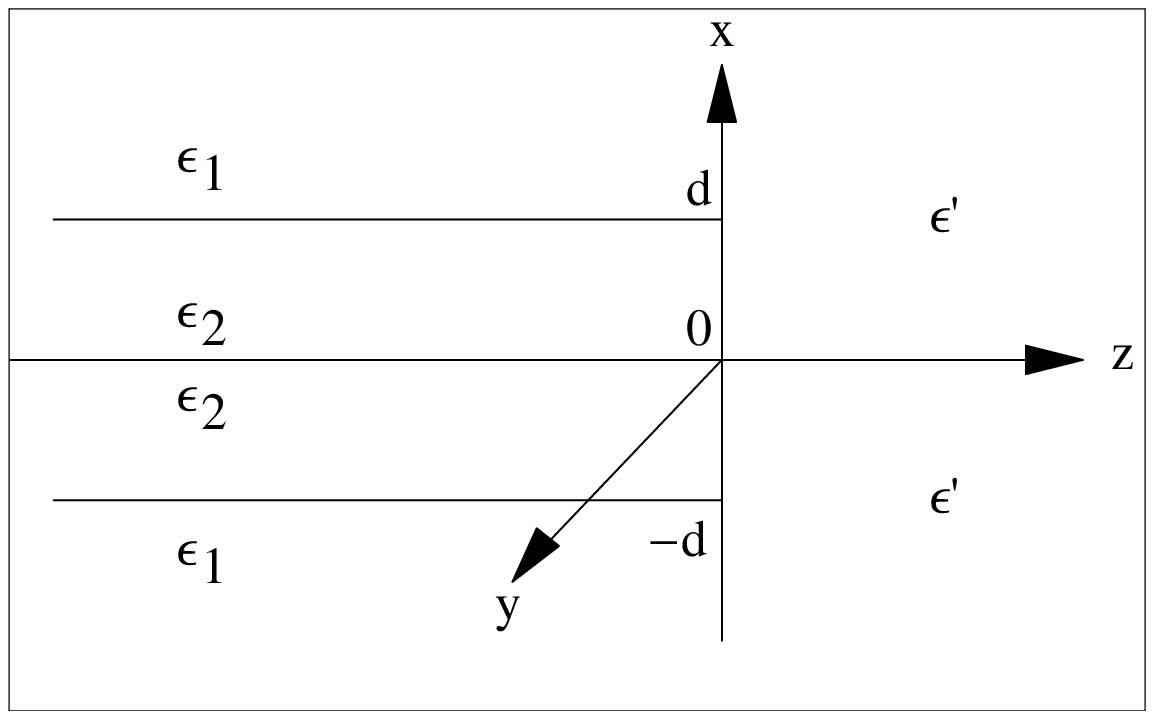}

  \caption{}
\end{figure}
\newpage
\clearpage
\newpage
\setlength{\unitlength}{1cm}
\begin{figure}
 \includegraphics{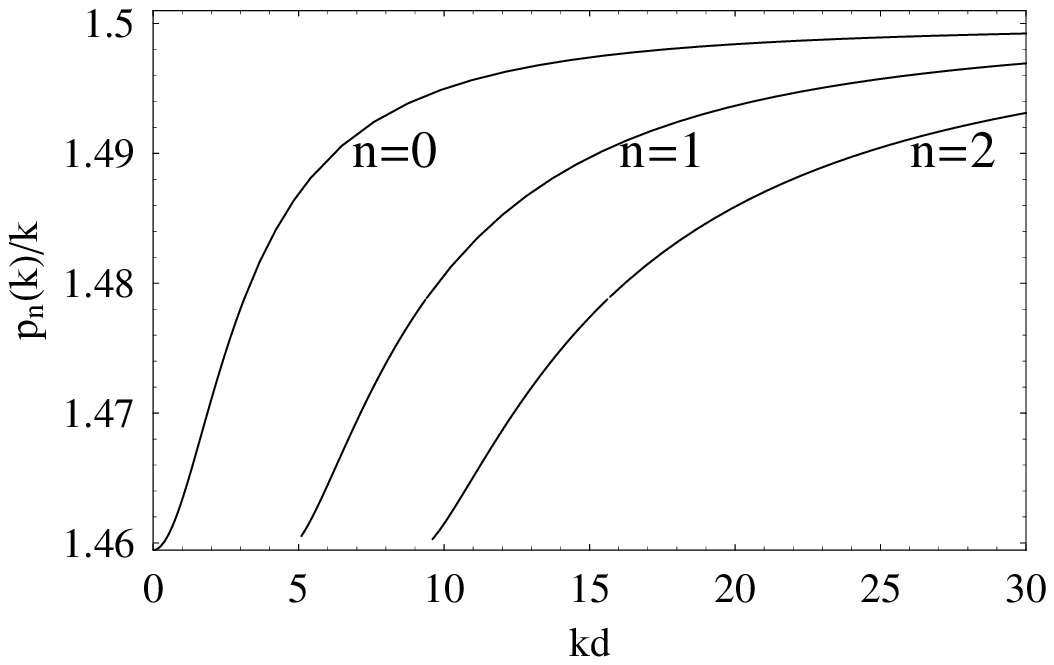}

  \caption{}
\end{figure}
\newpage
\clearpage
\newpage
\setlength{\unitlength}{1cm}
\begin{figure}
 \includegraphics{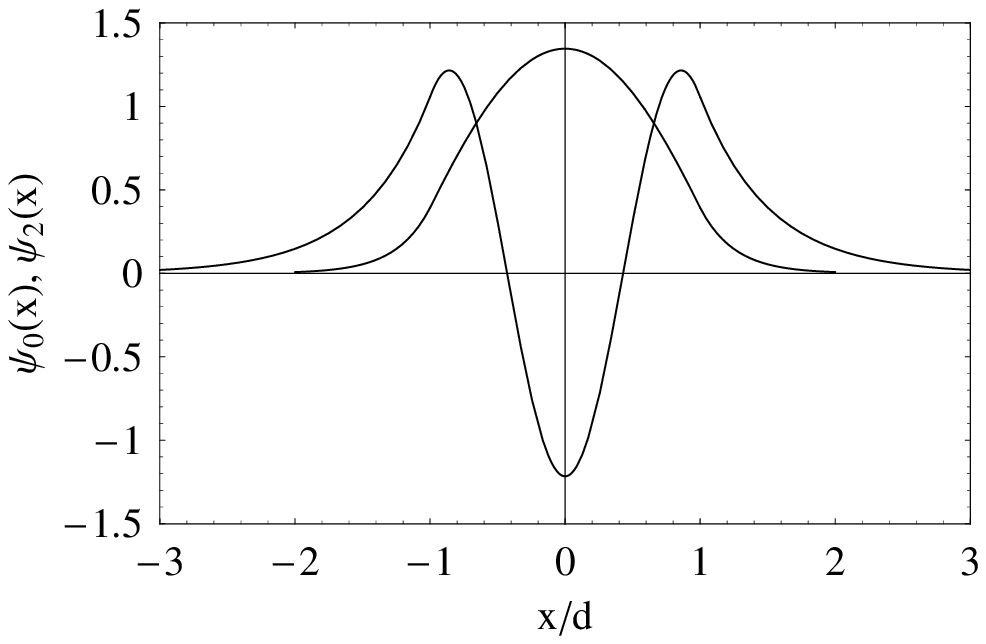}

  \caption{}
\end{figure}
\newpage
\clearpage
\newpage
\setlength{\unitlength}{1cm}
\begin{figure}
 \includegraphics{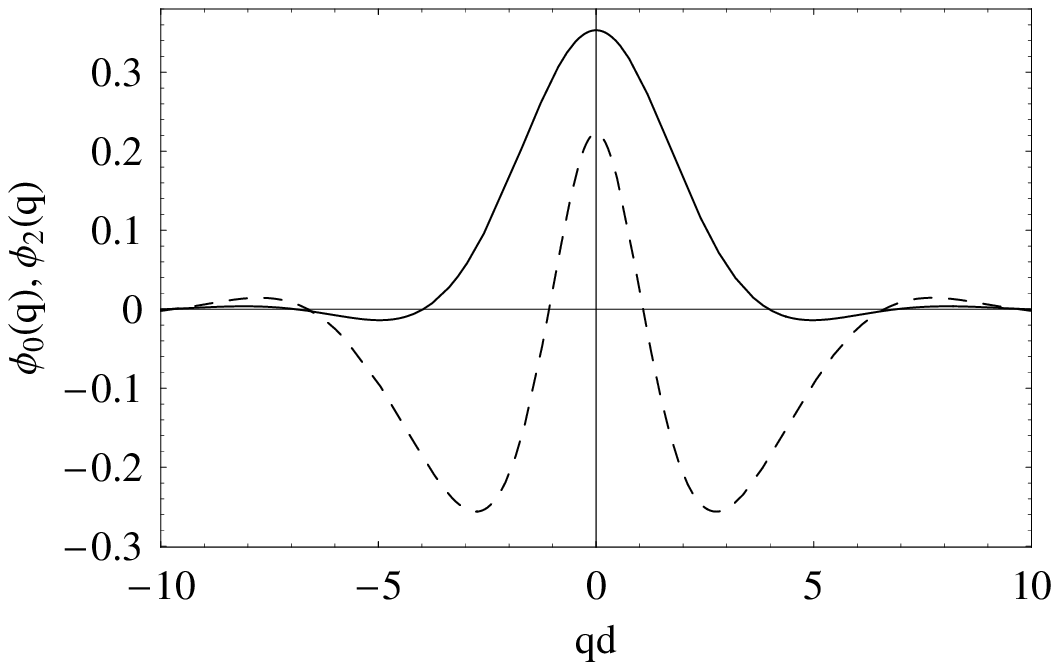}

  \caption{}
\end{figure}
\newpage
\clearpage
\newpage
\setlength{\unitlength}{1cm}
\begin{figure}
 \includegraphics{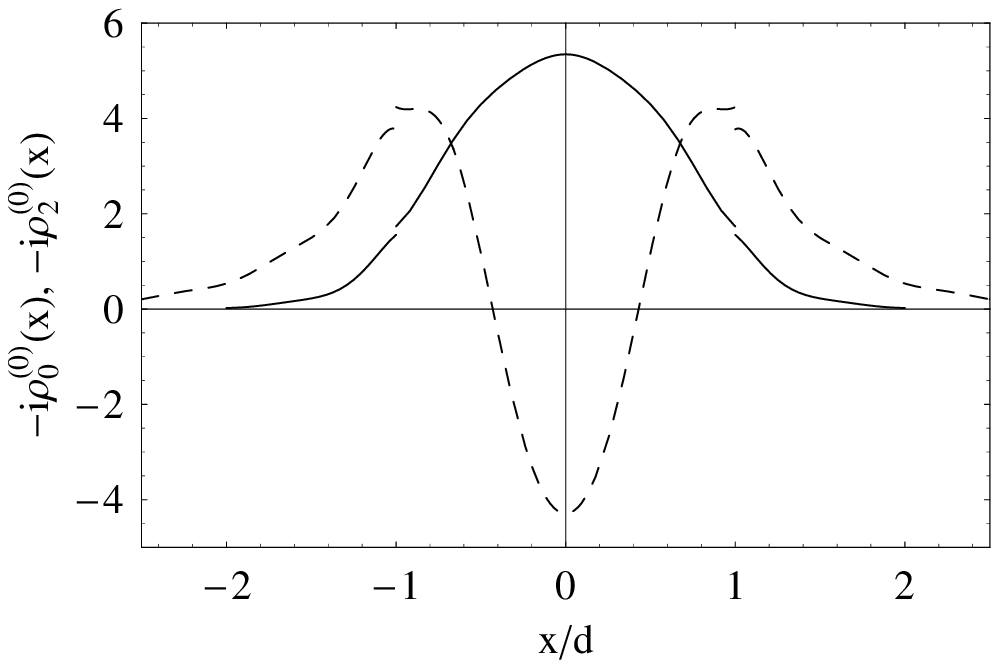}

  \caption{}
\end{figure}
\newpage
\clearpage
\newpage
\setlength{\unitlength}{1cm}
\begin{figure}
 \includegraphics{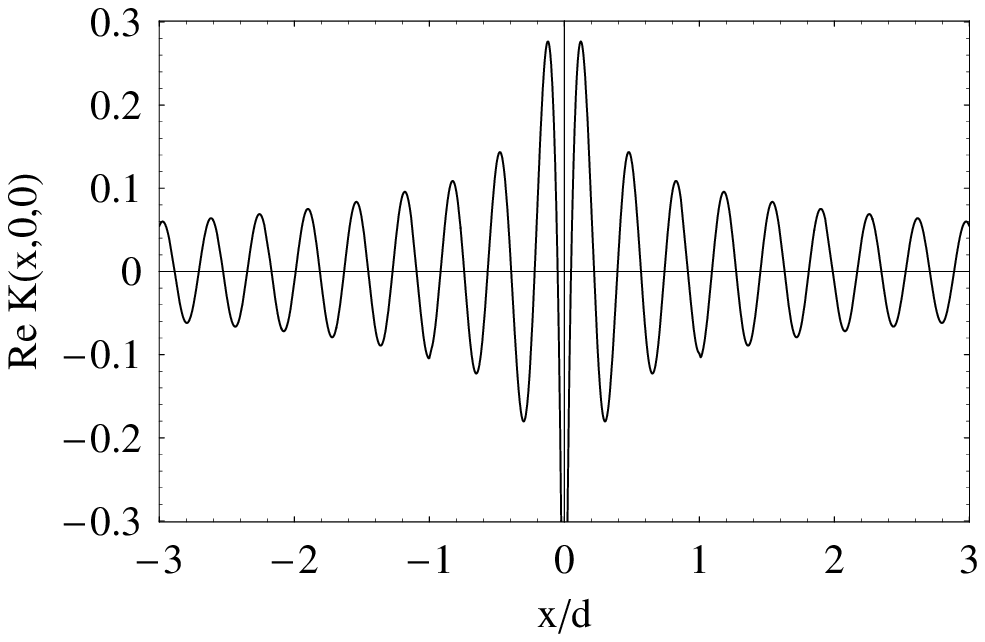}

  \caption{}
\end{figure}
\newpage
\clearpage
\newpage
\setlength{\unitlength}{1cm}
\begin{figure}
 \includegraphics{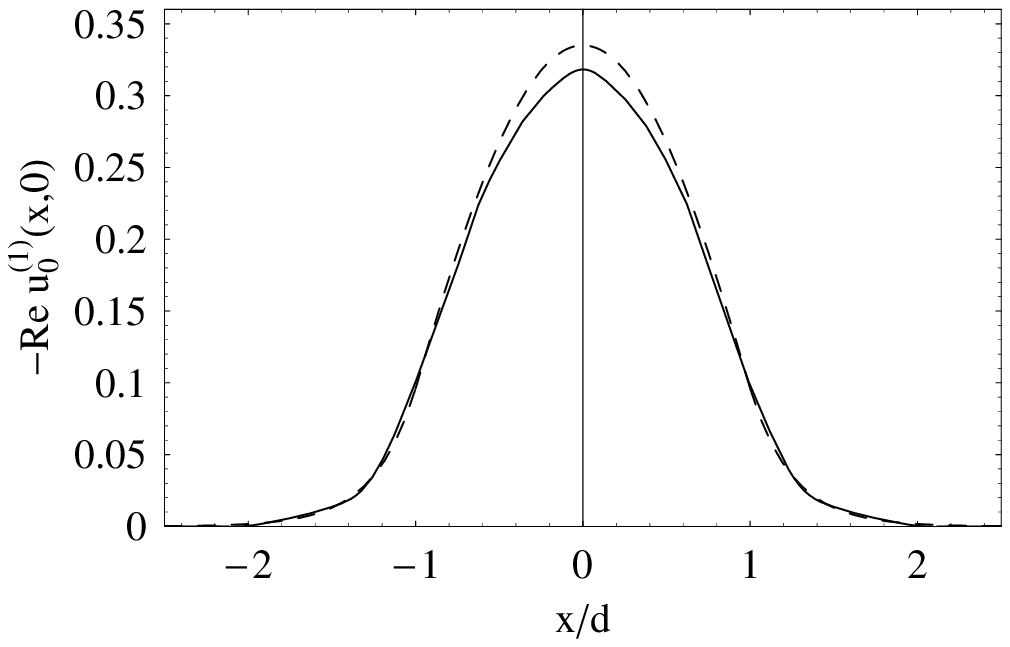}

  \caption{}
\end{figure}
\newpage
\clearpage
\newpage
\setlength{\unitlength}{1cm}
\begin{figure}
 \includegraphics{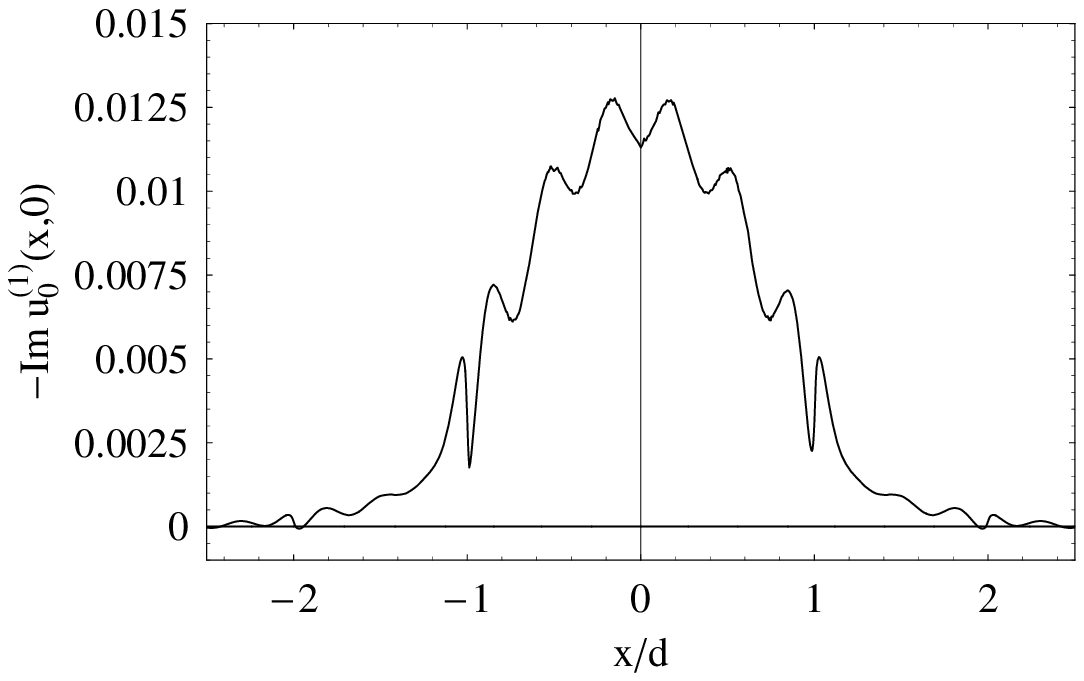}

  \caption{}
\end{figure}
\newpage
\clearpage
\newpage
\setlength{\unitlength}{1cm}
\begin{figure}
 \includegraphics{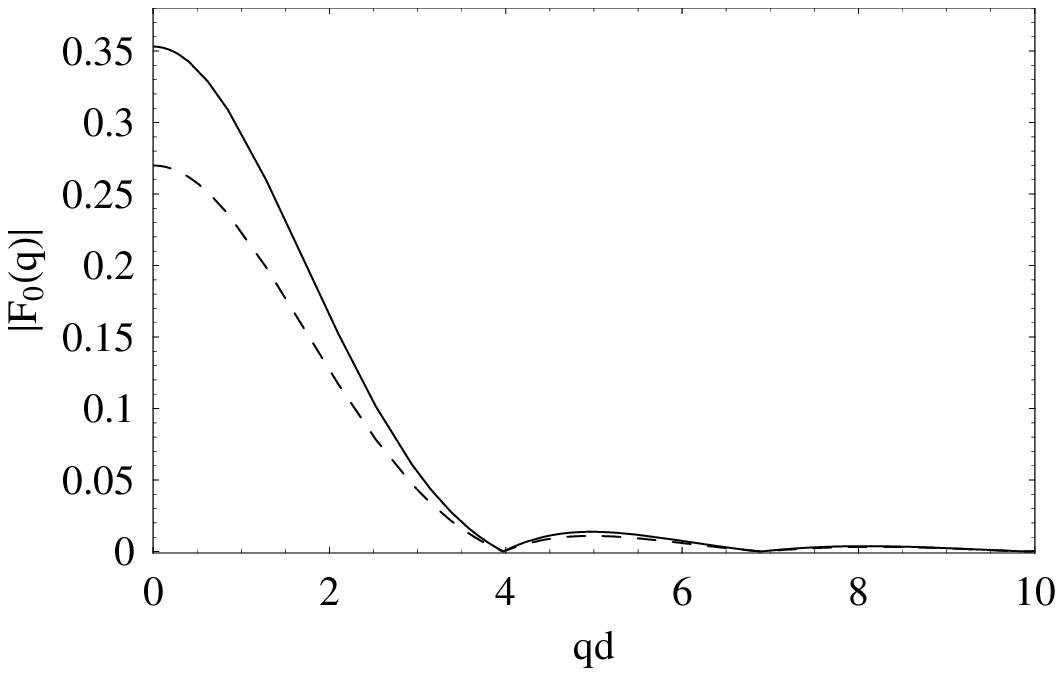}

  \caption{}
\end{figure}


\newpage
\begin{thebibliography}{99}

\bibitem{1}
H. Levine and J. Schwinger, Phys. rev. \vol{73}, 383 (1948).

\bibitem{2}
H. Levine and J. Schwinger, Comm. Pure Appl. Math. \vol{3}, 355 (1950).

\bibitem{3}
A. Sommerfeld, Math. Ann. \vol{47}, 317 (1896).

\bibitem{4}
A. Sommerfeld, {\it Optics} (Akademische Verlagsgesellschaft, Leipzig, 1964).

\bibitem{5}
H. Bouwkamp, Rep. Prog. Phys. \vol{17}, 35 (1954).

\bibitem{6}
M. Born and E. Wolf, {\it Principles of Optics} (Pergamon Press, Oxford, 1975).

\bibitem{7}
F. E. Gardiol in {\it Advances in electronics and electron physics}, edited by P. W. Hawkes, \vol{63}, 139 (1985).

\bibitem{8}
D. Hondros and P. Debye, Ann. d. Phys. \vol{32}, 465 (1910).

\bibitem{9}
G. F. Homicz and J. A. Lordi, J. Sound Vib. \vol{41}, 283 (1975).

\bibitem{9A}
P. Joseph and C. L. Morfey, J. Acoust. Soc. Am. \vol{105}, 2590 (1999).

\bibitem{10}
M. S. Howe, {\it Hydrodynamics and Sound} (Cambridge University Press, Cambridge, 2007).

\bibitem{11}
S. W. Rienstra and A. Hirschberg, {\it An Introduction to Acoustics} unpublished, available via internet (2013).

\bibitem{12}
D. M. S. Soh, J. Nilsson, S. Baek, C. Codemard, Y. Jeong, and V. Philippov, J. Opt. Soc. Am. \vol{A21}, 1241 (2004).

\bibitem{13}
A. E. Heins, Quart. Appl. Math. \vol{6}, 157 (1948).

\bibitem{13A}
P. C. Clemmow, Proc. Roy. Soc. A \vol{205}, 286 (1951).

\bibitem{14}
F. E. Borgnis and C. H. Papas, {\it Electromagnetic Waveguides and Resonators} (Springer, Berlin, 1958).

\bibitem{15}
D. Marcuse, {\it Theory of Dielectric Optical Waveguides} (Academic, New York, 1974).

\bibitem{16}
H. Kogelnik, in {\it Integrated Optics}, Topics Appl. Phys. \vol{7}, ed. by T. Tamir (Springer, Berlin, 1979).

\bibitem{17}
P. Lorrain, D. R. Corson, and F. Lorrain, {\it Electromagnetic Fields and Waves} (Freeman, New York, 1988).

\bibitem{18}
A. Bratz, B. U. Felderhof, and G. Marowsky, Appl. Phys. B \vol{50}, 393 (1990).

\bibitem{19}
R. G. Newton, {\it Scattering theory of waves and particles} (McGraw-Hill, New York, 1966).

\bibitem{20}
A. R. da Silva and G. P. Scavone, J. Phys. A \vol{40}, 397 (2007).

\bibitem{21}
W. Duan and R. Kirby, J. Acoust. Soc. Am. \vol{131}, 3638 (2012).



\end{thebibliography}
\end{document}